\documentclass{mem}
\usepackage{natbib}\usepackage{txfonts}\usepackage{balance}
\usepackage{graphicx}
\usepackage[a4paper,breaklinks,dvipdfm]{hyperref}
\idline{75}{282}
\begin{document}
\def\teff{$T\rm_{eff }$}
\def\kms{$\mathrm {km s}^{-1}$}

\title{
Star Clusters and Super Massive Black Holes: High Velocity Stars Production}

   \subtitle{}

\author{
G. \,Fragione\inst{1} 
\and R. \, Capuzzo-Dolcetta\inst{1}
          }

\institute{Department of Physics -- 
Sapienza University of Rome --
Piazzale Aldo Moro, 2, I-00185 Roma, Italy
\email{giacomo.fragione@uniroma1.it}
}

\authorrunning{Fragione}

\titlerunning{High Velocity Stars from Star Clusters}

\abstract{
One possible origin of high velocity stars in the Galaxy is that they are the product of the interaction of binary systems and supermassive black holes. We investigate a new production channel of high velocity stars as due to the close interaction between a star cluster and supermassive black holes in galactic centres. The high velocity acquired by some stars of the cluster comes from combined effect of extraction of their gravitational binding energy and from the slingshot due to the interaction with the black holes. Stars could reach a velocity sufficient to travel in the halo and even overcome the galactic potential well, while some of them are just stripped from the cluster and start orbiting around the galactic centre.
}
\maketitle{}

\section{Introduction}

High Velocity Stars (HVSs) are Galactic halo stars with high peculiar motions that can be divided into two groups by means of their origin and peculiar velocities. Runaway Stars (RSs) are Galactic halo stars with peculiar motions higher than $40$ km s$^{-1}$, but below the galactic escape velocity. RSs are produced in binary systems thanks to the velocity kick due to the supernova explosion of the former companion or dynamical multi-body interactions \citep{sin11}. Hyper Velocity Stars (HyVSs) are stars escaping the Galaxy with velocities up to over $1000$ km s$^{-1}$. Hills first predicted their existence as consequence of the tidal breakup of a binary passing close to a Black Hole (BH). Other mechanisms have been proposed to explain HVSs, as the interaction of a Black Hole Binary (BHB) with a single star or the arrival from another nearby galaxy \citep{brw15}.

By studying HyVSs, it is possible to infer information about different branches of physics, as the Galaxy gravitational potential and its Dark Matter component \citep{fra16c}.

\section{HVSs from Star Clusters}

Many galaxies show nucleated central regions, the so-called Nuclear Star Clusters (NSCs) \citep{bok09}. One mechanism suggests that massive stellar clusters, such as Globular Clusters (GCs), could spiral into the center of the galaxy where they merge to form a dense nucleus. In the infall scenario, due to strong interactions with the central massive BH or BHB, some stars can be accelerated to high velocities \citep{acs16}.

\subsection{Globular Clusters}

\begin{figure}[]
\resizebox{\hsize}{!}{\includegraphics[clip=true]{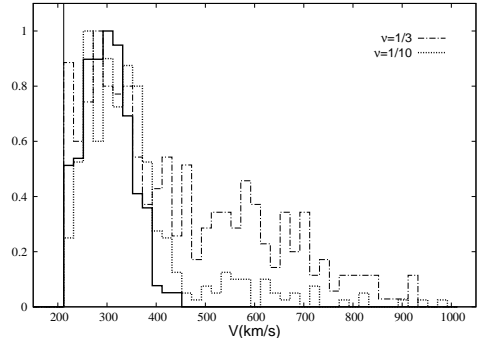}}
\caption{
\footnotesize
Velocity distributions of high velocity stars after the interaction between a GC and BH or BHB.}
\label{bhb}
\end{figure}

Our scattering experiments refer to the interaction of a super massive BH, a GC and a star. In our simulations, the black hole sits initially in the origin of the reference frame, while the GC follows same energy elliptical orbits with close pericentres \citep{cap15}. The branching ratio (BR) of stars, which leave the system becoming high velocity stars, is function of the GC orbital eccentricity and mass. The fraction of ejected stars decreases for less eccentric orbits and increases for larger GC masses, since the GC gravitational potential is more intense and is able to accelerate stars up to velocities high enough to escape the whole system. In the case of a BHB, the results of the scattering experiments depend on the BHB total mass $M$ and mass ratio $\nu=m_2/M$ (where $m_2$ is the mass of the lighter BH), the inclination of GC orbit and its mass \citep{fra16a}. However, while the branching ratio remains nearly constant at the value of the single BH for $\nu \lesssim 1/20$, for higher values it is an increasing function of $\nu$.  The velocity distributions of high velocity stars, in the case M$_{GC}=10^6$ M$_{\odot}$ and $M_{BHB}=10^8$ M$_{\odot}$, present a tail of HyVSs, which depends on the BHB mass ratio, with respect to the case of single BH (solid black line). For comparable BHs masses, as for $\nu=1/3$, such tail is huge, while it is negligible when one BH dominates, as for $\nu=1/10$. Stars with velocities $\gtrsim 500$-$700$ km s$^{-1}$, depending on the host galaxy mass and if it is spiral or elliptical, are able to escape the galaxy as HyVSs \ref{bhb}.

\subsection{Young Stellar Clusters}

The infall of a binary-rich young cluster (M$_{YC}=10^4$ M$_{\odot}$), starting from a distance of $100$ pc in elliptical orbit around the Milky Way's BH, is able to generate jets of HVSs. The Galactic potential is made up of a BH ($M_{BH}=4 \cdot 10^6$ M$_{\odot}$), a bulge ($M_{Bul}=3.76 \cdot 10^{9}$ M$_{\odot}$, $a=0.10$ pc), a disk ($M_{D}=5.36 \cdot 10^{10}$ M$_{\odot}$, $a=2.75$ pc, $b=0.30$ pc) and a logarithmic dark halo ($v_h=235$ km s$^{-1}$ at $R=8.15$ kpc). The cluster has a Kroupa initial mass function (2001) and Sana et al. (2012) initial period distribution, with $r_h=0.30$ pc [\citep{ohk15}]. About $0.1\%$ of stars become HyVSs, some of which are binaries. Some of these binaries will merge and produce blue stragglers HyVSs [\citep{fra16b}].

\section{Conclusions}

Our studies suggest that the interaction of star clusters with BHs could produce HVSs, some of which are able to escape the host galaxy. The results depend on the the cluster pericentre, while the presence of a BHB enhances the loss of stars and HyVSs. Finally, information on star clusters progenitors can be obtained by studying these clusters of escaped stars.

\bibliographystyle{aa}

\end{document}